\documentclass[aps,prb,twocolumn]{revtex4}
\usepackage{graphicx,epsf}

\newcommand{\ybco}{YBa$_2$Cu$_3$O$_{6+y}$}

\newcommand{\bscco}{Bi$_2$Sr$_2$CaCu$_2$O$_{8+\delta}$}
\newcommand{\lbco}{La$_{2-x}$Ba$_x$CuO$_4$}

\begin{document}

\title {Phenomenological lattice model for dynamic spin\\
and charge fluctuations in the cuprates}

\author{Matthias Vojta}
\affiliation{\mbox{Institut f\"ur Theorie der Kondensierten
Materie, Universit\"at Karlsruhe, 76128 Karlsruhe, Germany.}}

\author{Subir Sachdev}
\affiliation{Department of Physics, Yale University, P.O. Box
208120, New Haven CT 06520-8120, USA.}

\date{August 20, 2004}

\begin{abstract}
Motivated by recent neutron scattering experiments on the cuprate
superconductors, we present a phenomenological framework
describing the dynamics of collective spin excitations coupled to
charge/bond order fluctuations. Our quantum lattice model contains
two order parameter fields, and can capture spin excitations both
in broken-symmetry states with static lattice modulations, as well
as in homogeneous states where the charge/bond order is
fluctuating. We present results for different types of static
charge/bond order, namely site- and bond-centered stripes, and
plaquette modulation.
\end{abstract}
\pacs{74.72.-h,75.10.Jm}

\maketitle


\section{Introduction}

For certain cuprate high-temperature superconductors it has been established
that incommensurate spin and charge correlations,
commonly referred to as stripes, appear over a
significant range of the phase diagram \cite{lsco,mook,waki}.
The role of these stripes both for superconductivity itself as well as for
various anomalies in the normal state has been discussed
extensively \cite{ssrmp,stevek,ek,pnas,jan,doug,bondsite}, but is at present
not fully understood.

Recent neutron scattering experiments
\cite{jt04,hinkov,buyers,hayden} have mapped out spin excitations
in various cuprates over a large range of energies. Tranquada {\em
et al.} \cite{jt04} investigated the excitation branches in \lbco\
at a hole doping of $x=\frac 1 8$, which displays static spin and
charge order, up to energies of 200 meV. They found the
high-energy part to be well described by the spectrum of spin
ladders, pointing towards bond order in the material
\cite{MVTU,GSU}. Remarkably, other cuprate families like \ybco\ or
\bscco, where charge order remains dynamically fluctuating, show a
very similar spin excitation spectrum at elevated energies. (At
low energies, the spectrum depends on whether the system shows
static order; furthermore, the strength of electronic
quasiparticle damping is different between different compounds,
limiting the wavevector range where sharp single-particle
excitations can be seen.) These experiments raise the question of
whether there exists a dynamic spin response which is {\em
universal} among all cuprates: namely a high-intensity
(``resonance'') peak at wavevector $(\pi,\pi)$ and energy between
20 and 50 meV, with both downward and upward dispersing branches
of excitations.

The purpose of this paper is to develop a unified theoretical
framework for describing spin excitations in the presence of both
static and dynamic charge order. We will do this using a
phenomenological lattice model. Our approach differs from previous
phenomenological theories \cite{zachar,balatsky,zds} in two
important ways: ({\em i\/}) These theories took the continuum
limit for spin modulations in the vicinity of the incommensurate
spin-ordering wavevector; such a theory is not expected to be
valid near the resonance peak at $(\pi,\pi)$. We will instead use
a lattice model which is valid both at $(\pi, \pi)$ and near the
incommensurate wavevector, and for a wide range of energies. ({\em
ii\/}) Our model explicitly selects {\sl collinear} spin
correlations, in contrast to the previous approaches
\cite{zachar,zds} where the distinction between collinear and
spiral spin correlations only appears in higher order terms which
are not fully accounted for in existing calculations.

Our model is an extension of recent microscopic spin-only models
for stripe phases \cite{MVTU,GSU} which have quite successfully
modelled some of the neutron scattering data.\cite{jt04} One of
our purposes is to judge the significance of this agreement
between theory and experiment: in particular, we wish to delineate
the range of models which are compatible with the data, including
those which cannot be described by simple spin-only Hamiltonians.
Furthermore, our approach can be naturally extended to the case
where the charge order is dynamic, rather than the static order
needed to define the models of Refs.~\onlinecite{MVTU,GSU}.

We will model spin and charge fluctuations on a phenomenological
basis using Landau order parameters for both. The coupling between
the two orders can shift the minimum energy of the spin
excitations from $(\pi,\pi)$ to the incommensurate wavevector
dictated by the charge order. On a microscopic scale, the
influence of the charge order on the spin sector can be understood
in terms of both spatially modulated spin densities and spatially
modulated couplings, as found in the models of
Refs.~\onlinecite{MVTU,GSU}. Our Landau-like theory will be
formulated directly on the underlying square lattice, which will
allow us to capture lattice effects such as the differences
between site-centered or bond-centered order. Importantly, the
spin sector described by our theory will be strongly fluctuating,
{\em i.e.\/}, we are far from the semiclassical limit described by
spin waves \cite{scheidl,erica,balatsky}. Microscopically, these
fluctuations can arise from the tendency to dimerization, {\em
i.e.\/}, from {\sl bond order}, which is present in the undoped
paramagnetic parent Mott insulator \cite{ssrmp,vs}. (Note that
bond order can occur both in site-centered and bond-centered
stripe states.)


\section{Quantum lattice model}

We assume a dominant antiferromagnetic interaction between the
spins, and so  model the quantum spin fluctuations by a standard
vector $\varphi^4$ Landau theory for the antiferromagnetic order
parameter at the {\em commensurate\/} wavevector ${\bf Q} = (\pi,
\pi)$. So on a square lattice of sites, $j$, we parameterize the
lattice spins by
\begin{equation}
S_{j\alpha} \propto e^{i {\bf Q} \cdot {\bf r}_j} \varphi_{j
\alpha}
\end{equation}
where $\alpha=x,y,z$. In the absence of any coupling to
charge/bond order, we assume that the dominant spin fluctuations
remain at the commensurate ${\bf Q}$; we find below that this
feature is important in obtaining a resonance peak at ${\bf Q}$.
The effective action for these commensurate spin fluctuations has
a familiar form:
\begin{eqnarray}
\mathcal{S}_0 &=& \int d \tau \sum_j \left[ \frac{1}{2} \left(
\frac{\partial \varphi_{j\alpha}}{\partial \tau} \right)^2 +
\frac{s}{2} \varphi_{j\alpha}^2 + \frac{u}{4} \left(
\varphi_{j\alpha}^2\right)^2 \right] \nonumber \\
&+& \int d \tau \sum_{\langle j j' \rangle} \frac{c^2}{2}
\left(\varphi_{j\alpha} - \varphi_{j' \alpha} \right)^2
\end{eqnarray}
In principle, there should also be Berry phases in the quantum
spin action, but we assume that they have averaged out to zero:
this is expected to be valid in the compressible superconducting
states, or in the incompressible Mott insulators with an even
number of electrons per unit cell, but likely not at the quantum
critical point between such phases.

Now we include the effect of charge/bond order. This we represent
by the complex continuum fields $\phi_{x,y} ({\bf r}, \tau)$ which
measure the amplitude of charge order at the wavevectors ${\bf
K}_x = (\pi/2, 0)$ and ${\bf K}_y = (0, \pi/2)$ -- this is the
dominant ordering wavevector of the Mott insulating state at 1/8
doping, and is therefore the appropriate reference wavevector for
our considerations: we will show below how deviations in the
charging ordering wavevector from ${\bf K}_{x,y}$ in the
superconducting phases (or at non-zero temperatures) can be easily
built into our formalism. Using the complex order parameters
$\phi_{x,y}$, it is convenient to define the real field
\begin{equation}
Q_x ({\bf r} ) = \phi_x ({\bf r}) e^{i {\bf K}_x \cdot {\bf r}} +
\phi_x^{\ast} ({\bf r}) e^{-i {\bf K}_x \cdot {\bf r}}
\end{equation}
and similarly for $Q_y$. For ${\bf r}$ on the sites of the square
lattice, the $Q_{x,y}$ are measures of the charge density
modulation on those sites. On the other hand, for ${\bf r}$ on the
links of the square lattice, the $Q_{x,y}$ is a measure of the
local {\sl bond} order: this is determined by the modulation in
the local pairing amplitude or exchange energy. With these
physical interpretations at hand, we can write down the following
couplings between the spin and charge fluctuations
\begin{eqnarray}
&& \!\!\!\!\!\!\!\!\! \mathcal{S}_x = \int d \tau \sum_j \Bigl[
\lambda_1 Q_x ({\bf r}_j ) \varphi_{j\alpha}^2 + \lambda_2 Q_x
({\bf r}_{j+x/2})
\varphi_{j\alpha} \varphi_{j+x,\alpha} \nonumber \\
&& \!\!\!\!\!\!\!\!\! + \lambda_3 Q_x ({\bf r}_j)
\varphi_{j-x,\alpha} \varphi_{j+x,\alpha} + \lambda_4 Q_x ({\bf
r}_{j+y/2}) \varphi_{j\alpha} \varphi_{j+y,\alpha} \Bigr]
\end{eqnarray}
with four independent couplings constants $\lambda_{1-4}$; the
same couplings will appear in the corresponding $\mathcal{S}_y$.
Notice that $\lambda_1$ implements the correlation between the
on-site charge density and the amplitude of the spin fluctuations,
while $\lambda_{2-4}$ ensure that the effective first- and
second-neighbor exchange constants controlling the spin
correlations modulate along with the bond order.

If we now take $\phi_{x,y}$= constant, then the action
$\mathcal{S}_0 + \mathcal{S}_{x,y}$ represents our general theory
for quantum spin fluctuations in a background of static
charge/bond order. The site-centered case has $\phi_x = 1$, and
the bond-centered case has $\phi_x = e^{i \pi/4}$. Two-dimensional
(plaquette or checkerboard) order will have both $\phi_x$ and
$\phi_y$ non-zero. At the Gaussian level ($u=0$), the problem is
quadratic in the $\varphi$ fields, and can be solved by
diagonalizing a matrix of size $(N_x N_y)^2$, where $N_{x,y}$
describe the size of the unit cell (in the charge sector).

For large enough $\lambda$ couplings, the minimum energy of the
$\varphi$ fluctuations will be shifted away from $(\pi,\pi)$, as
observed in experiment. Notably, the restriction to {\em real}
$\varphi$ implies that the spin order remains collinear. For small
mass $s$, the spin order can condense as usual, and fluctuations
around the condensate will lead to low-energy Goldstone modes.

\subsection{Fluctuating charge order}

The above formalism is designed to allow easy extension to the
dynamic charge order case, which is likely relevant for \ybco\ and
\bscco. We use a continuum formulation to describe the charge
fluctuations, with the following general action consistent with
all the symmetries of the lattice:
\begin{eqnarray}
&&  \mathcal{S}_{\phi} = \int d \tau d^2 {\bf r} \Bigl[ \left|
\partial_\tau \phi_x \right|^2 + \left|
\partial_\tau \phi_y \right|^2 + c_1^2 \left|
\partial_x \phi_x \right|^2 \nonumber \\
&& \!\!\!\!\!\!\! + c_2^2 \left|
\partial_y \phi_x \right|^2 + c_1^2 \left|
\partial_y \phi_y \right|^2 + c_2^2 \left|
\partial_x \phi_y \right|^2 + i \delta \phi_x^{\ast} \partial_x \phi_x
\nonumber \\
&& \!\!\!\!\!\!\! + i \delta \phi_y^{\ast}
\partial_y \phi_y + s_1 \left( |\phi_x|^2 + |\phi_y|^2 \right) +
u_1 \left(|\phi_x|^4 + |\phi_y|^4 \right)
\nonumber \\
&&+ v |\phi_x|^2 |\phi_y|^2 + w \left( \phi_x^4 + \phi_x^{\ast 4}
+ \phi_y^4 + \phi_y^{\ast 4} \right) \Bigr]
\end{eqnarray}
Note especially the term proportional to $\delta$: it is
generically present, and when the strength of the charge order is
weak it ensures that the dominant charge order fluctuations are at
an incommensurate wavevector unequal to ${\bf K}_{x,y}$. However
at low temperatures ($T$) in an insulating state, the `lock-in'
term proportional to $w$ eventually dominates, and selects a
commensurate charge-ordered state with wavevectors equal to ${\bf
K}_{x,y}$. The quartic $v$ term determines whether the long-range
order will be one-dimensional (``stripe'') or two-dimensional
(``plaquette'', ``checkerboard'').

The effect of charge fluctuations on the spin excitation spectrum
can now be determined by self-consistently computing the frequency
and momentum dependence of the $\varphi_{\alpha}$ self energy to
second order in the $\lambda$'s, similar to
Ref.~\onlinecite{kampf}. A full calculation along this line is
beyond the scope of this paper, we expect that in the limit of
small mass $s_1$ the results will be similar to the ones obtained
in the static $\phi$ theory.


\subsection{Coupling to phase fluctuations}

An advantage of the present phenomenological formalism is that is
allows easy extension to include couplings to other collective
modes. If the cuprate compound is a superconductor, then there is
an additional mode associate the fluctuation of $\theta$, the
phase of the superconducting order. As in Ref.~\onlinecite{frey},
the most relevant coupling of $\theta$ to the spin fluctuations is
\begin{equation}
\mathcal{S}_{\varphi\theta} = \int d^2 r d\tau \left[ i \gamma
\partial_\tau \theta \varphi_{\alpha}^2 \right]
\label{theta1}
\end{equation}
The action for $\theta$ fluctuations can be generally written as
\cite{mpaf}
\begin{equation}
\mathcal{S}_{\theta} = \int \frac{d^2 k d \omega}{8 \pi^3} (K_1
k^\sigma \omega^2 + K_2 k^2) |\theta(k, \omega)|^2 \label{theta2}
\end{equation}
where $k$ is a wavevector and $\omega$ is an imaginary frequency,
and the parameter $\sigma$ is determined by the nature of the
Coulomb interaction: for screened short-range interactions
$\sigma=0$, while for in-plane $1/r$ Coulomb interactions with
independent layers $\sigma =1$. In a paramagnetic state where
$\varphi_\alpha$ excitation forms a sharp $S=1$ `triplon'
excitation, the coupling to phase fluctuations will induce damping
in the triplon spectral function. At $T=0$, and at the bottom of
the triplon band, it is not difficult to compute from
Eqs.~(\ref{theta1},\ref{theta2}) that the imaginary part of the
triplon self energy is $\sim (\epsilon-\Delta)^{2(d-\sigma)/(2-\sigma)}$ at the bottom of
the band ($\epsilon \geq \Delta$ is real frequency and $\Delta$ is
the spin gap): this arises from the `radiation' of $\theta$
excitations by the triplon.


\section{Results for static charge order}

\begin{figure}[!t]
\includegraphics[width=3.1in]{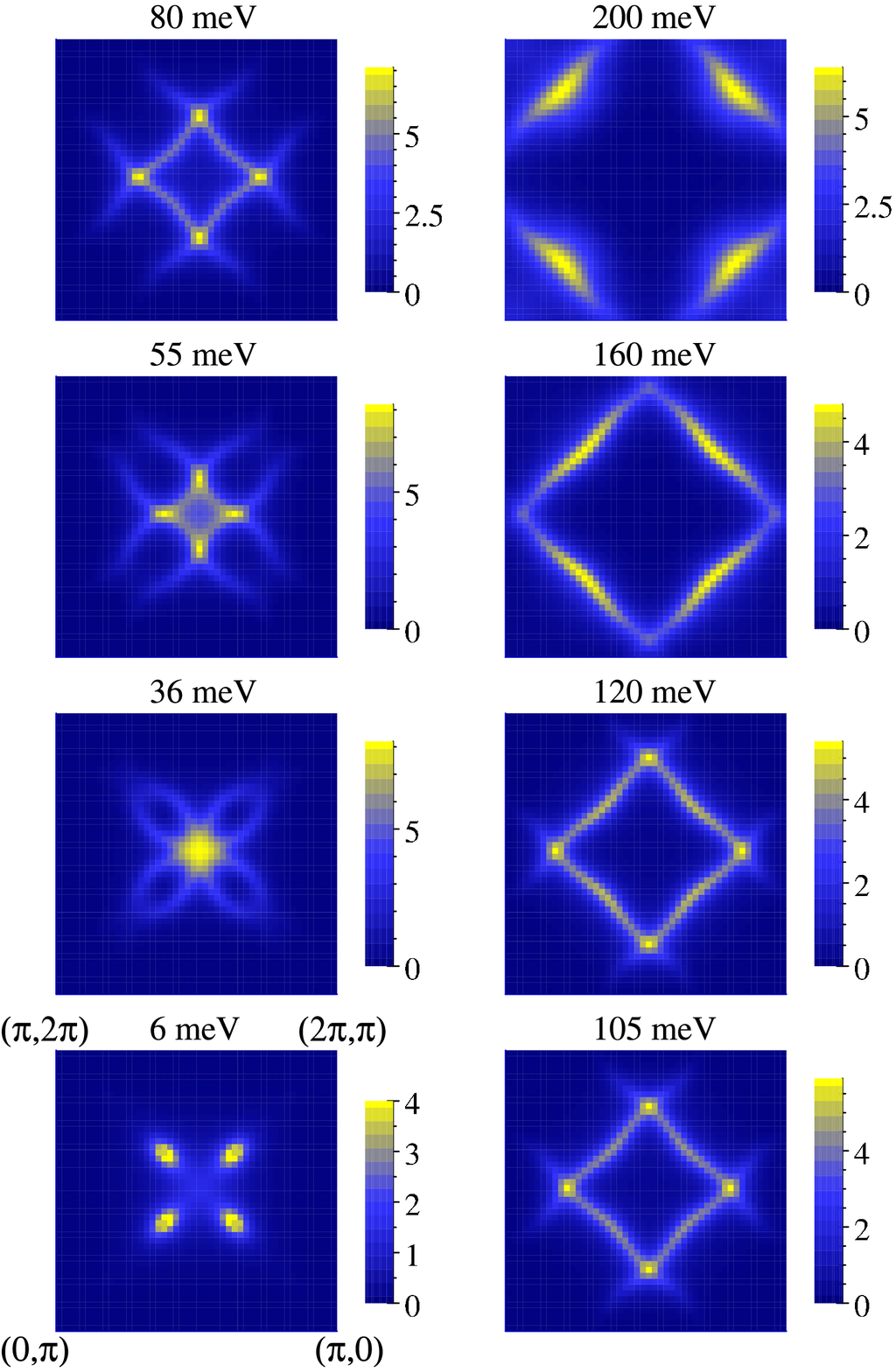}
\caption{(color)
Neutron scattering intensity, $\chi''({\bf k},\omega)$,
for a state with bond-centered charge order.
Parameter values are
$c=160$ meV, $\lambda_1 |\phi_x| =100$ meV, $\lambda_2 |\phi_x|= 600$ meV,
$\lambda_3=\lambda_4=0$.
$s=550$ meV is choosen is place the spin sector at its critical point.
The panels show fixed energy cuts as function of momentum in the
{\em magnetic} Brillouin zone.
The $\delta$ peaks have been replaced by Lorentzians with width $\Gamma = 15 meV$,
and the responses of horizontal and vertical stripes have been added.
The figure can be directly compared to Fig. 2 of Ref.~\protect\onlinecite{jt04}.
}
\label{fig:res1}
\end{figure}

\begin{figure}[!t]
\includegraphics[width=3.1in]{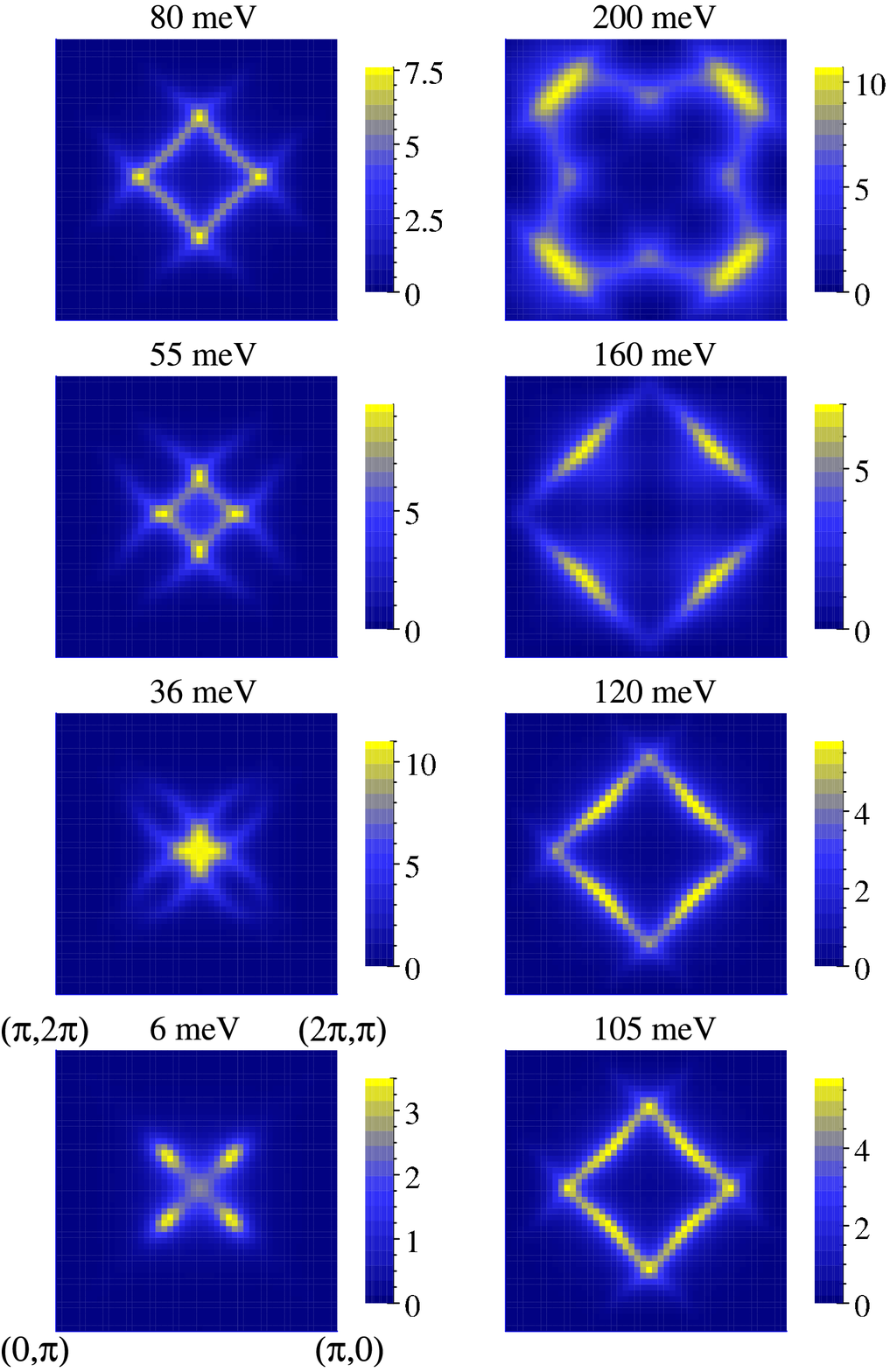}
\caption{(color)
As Fig.~\ref{fig:res1}, but now for a state with {\em site}-centered charge order,
and parameter values are
$\lambda_1 |\phi_x| =200$ meV, $\lambda_3 |\phi_x|= 100$ meV,
$\lambda_2=\lambda_4=0$, and $s=140$ meV.
Again the responses of horizontally and vertically ordered states have been added.
}
\label{fig:res2}
\end{figure}

\begin{figure}[!t]
\includegraphics[width=3.1in]{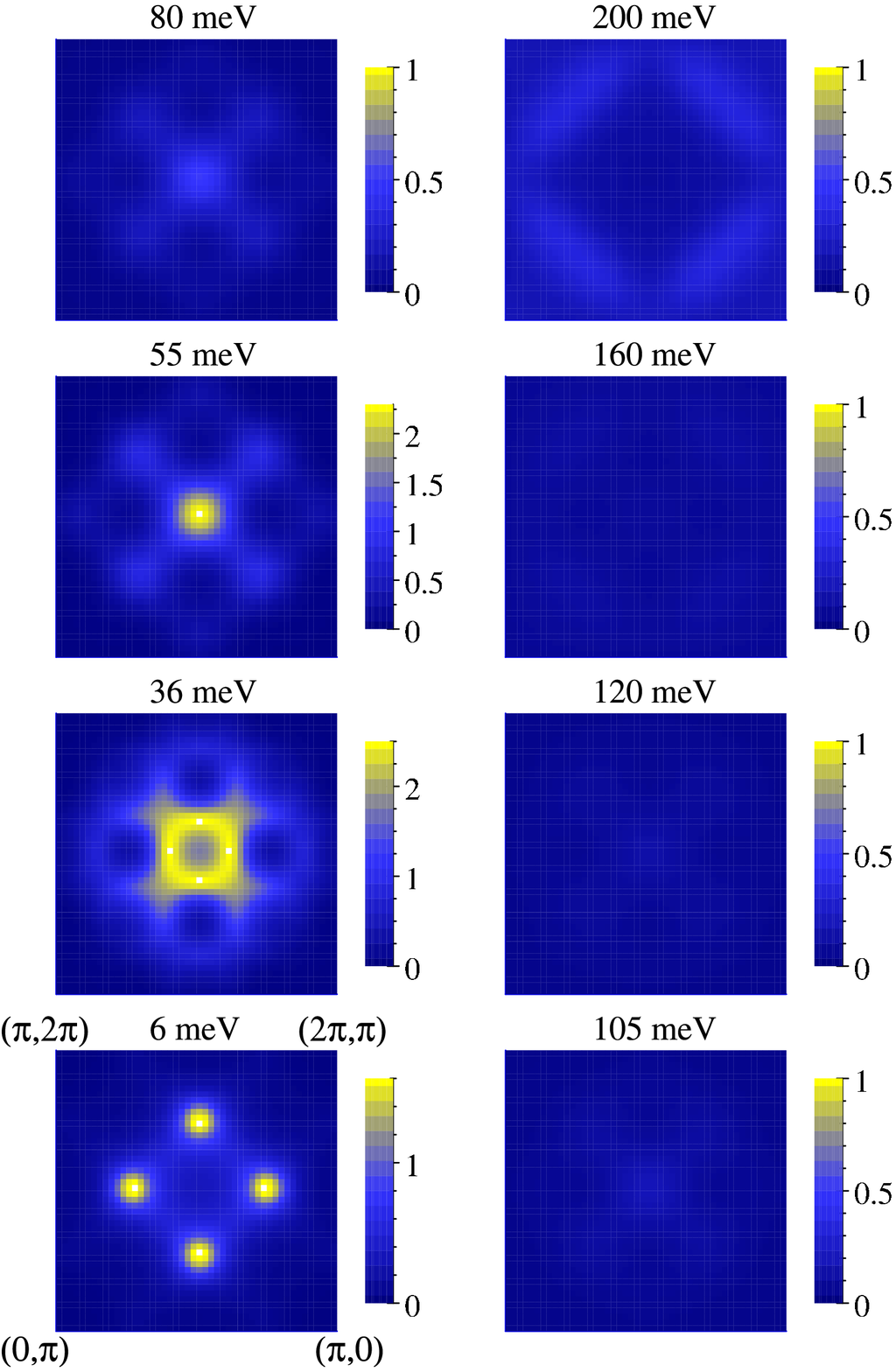}
\caption{(color)
As Fig.~\ref{fig:res1}, here for a state with two-dimensional {\em plaquette} charge order.
Parameters are
$\lambda_1 |\phi_x| =100$ meV, $\lambda_2 |\phi_x| = 600$ meV,
$\lambda_3=\lambda_4=0$, and $s=1.3$ eV.
The spin order condenses at wavevectors $(\pi\pm\pi/4,\pi\pm\pi/4)$ instead of $(\pi\pm\pi/4,\pi)$,
$(\pi,\pi\pm\pi/4)$.
Furthermore the dispersion features a single energy scale only, i.e., there is
no magnetic response above the saddle point at $(\pi,\pi)$.
(A higher band of spin excitations start around 250 meV.)
}
\label{fig:res3}
\end{figure}

Let us now present a few result for the case of static charge/bond
order, calculated in the Gaussian approximation of $\mathcal{S}_0
+ \mathcal{S}_x + \mathcal{S}_y$. We directly calculate the
$T\!=\!0$ susceptibility as measured in inelastic neutron
scattering. In the following, we restrict our attention to the
one-particle contributions, as the multiparticle continuum will be
hard to detect experimentally. Averaging over the neutron spin
polarizations, we obtain the dynamic spin susceptibility
$\chi''({\bf k},\omega)$ as sum of $\delta$ peaks with weights
determined by various matrix element terms.

In order to compare with the experiment of Ref.~\onlinecite{jt04}
we add the contributions from horizontal and vertical charge modulations
(Figs.~\ref{fig:res1} and \ref{fig:res2}),
and plot the result as function of the external momentum
at fixed energy, furthermore we broaden the
$\delta$ peaks to account for the experimental resolution.

In Figs.~\ref{fig:res1} and \ref{fig:res2} we show the response
for bond- and site-centered stripe structures, where the
parameters are chosen to match the experimental result of
Ref.~\onlinecite{jt04}. The coupling between charge and spin
sector is captured by $\lambda_1$ which modulates the spin
density, and in $\lambda_{2,3}$ which induce a spatial variation
in the exchange couplings. As in Refs.~\onlinecite{MVTU,GSU}, we
observe a ``dual'' character of the lowest spin excitation branch:
For small energies ({\em e.g.\/} 6 meV) the response consists of
four peaks representing the four cones of spin-wave modes. With
increasing energy the cones widen; however, the outer part becomes
suppressed in intensity due to matrix element effects. Around 30
-- 50 meV the spectrum is dominated by the strong response near
$(\pi,\pi)$ (``resonance peak''), which arises from a saddle point
of the mode dispersion. For higher energies, the modes gradually
change their character towards a one-dimensional excitation
spectrum, and the scattering intensity forms a diamond which moves
outward with increasing energy. Overall, there is reasonable agreement with the
experimental data of Tranquada {\em et al.} \cite{jt04}.
A crucial point is the presence of two energy scales
in the dispersion: a bandwidth of about 250 meV arising from the
strong coupling along the stripes, and a saddle point at about 40
meV whose energy is dictated by the coupling across the stripes
and the deviation of the ordering wavevector from $(\pi,\pi)$.
Remarkably, the difference between the bond- and site-centered
situation are minimal, i.e., the is little distinction on symmetry
grounds. (Microscopically, however, strong spin fluctuations which
drive the system away from the quasiclassical limit are favored in
a bond-centered geometry \cite{MVTU,GSU,anisimov,lorenzana}.)

Fig.~\ref{fig:res3} shows the result for bond-centered checkerboard order,
with microscopic parameters similar to Fig.~\ref{fig:res1}.
Clearly, the result is completely different:
The low-enery modes appear at four points rotated by 45 degrees compared
to Figs.~\ref{fig:res1}, \ref{fig:res2}, and
the dispersion shows only a single energy scale.
Thus, the ``dual'' character of the spectrum is absent here.


\section{Conclusions}

We have presented a general quantum lattice model which describes
spin excitations in the presence of either static or fluctuating
charge/bond order.

In the case of anisotropic (`one-dimensional') charge order, the
magnetic modes resemble semi-classical spin waves at low energies,
but cross over into triplon excitations of a quasi-one-dimensional
quantum paramagnet at higher energies. The crossover energy,
associated with a saddle point in the mode dispersion, is only a
fraction of the bandwidth for systems close to a magnetic quantum
phase transition. Two ingredients are crucial for the ``dual''
character of the modes: ({\em i\/}) strong quantum fluctuations,
and ({\em ii\/}) the presence of two energy scales in the
dispersion. We found that the spectrum was relatively insensitive
to the bond- or site-centered nature of the charge order. We also
examined fully two-dimensional `plaquette' ordered states, and
found that they could not describe the experimental observations.

An interesting open question is the precise link between the
neutron scattering observations and the modulations observed in
recent STM experiments.\cite{ali,kapitul,seamus,oxy} In principle,
our approach can be adapted to any specific charge order observed
in STM, and can then compute its spin excitation spectrum. It is
already clear from our results that at least in \lbco, the charge
order cannot be strictly a two-dimensional `checkerboard'
structure: anisotropic ``quasi-one-dimensional'' correlations over
some finite range appear to be required.


\acknowledgments

We thank S.~Kivelson, A. Polkovnikov, S.~Scheidl, J.~Tranquada,
G.~Uhrig, and T. Ulbricht for illuminating discussions and
collaborations on related work. This research was supported by the
National Science Foundation under grants DMR-0098226 (S.S.) and
DMR-0210790, PHY-9907949 at the Kavli Institute for Theoretical
Physics (M.V., S.S.) and the DFG Center for Functional Nano\-structures
Karls\-ruhe (M.V.).


\end{document}